 \documentclass[11pt,reqno]{amsart}
 \usepackage{amsxtra}
 \usepackage[innercaption]{sidecap}
 \usepackage{graphicx}
 \graphicspath{{./images/}}
 \usepackage{amssymb}
 \usepackage{enumitem}

\theoremstyle{plain}

\def\CC {{\mathbb C}}

\def\NN {{\mathbb N}}
\def\ZZ {{\mathbb Z}}
\def\PP {{\mathbb P}}

\def\be {\begin{eqnarray}}
\def\ben {\begin{eqnarray*}}
\def\ee {\end{eqnarray}}
\def\een {\end{eqnarray*}}

\def\AAA{\kern-0.3em}
\def\AA{\kern-0.18em}
\def\AC{\kern-0.14em}
\def\AB{\kern-0.22em}

\newcommand \nc {\newcommand}

\newtheorem{theorem}{Theorem}[section]
\newtheorem{lemma}[theorem]{Lemma}
\newtheorem{proposition}[theorem]{Proposition}
\newtheorem{corollary}[theorem]{Corollary}
\newtheorem{definition}[theorem]{Definition}
\newtheorem{example}[theorem]{Example}
\newtheorem{remark}[theorem]{Remark}
\nc \bth[1] { \begin{theorem}\label{t#1} } \nc \ble[1] {
\begin{lemma}\label{l#1} } \nc \bpr[1] {
\begin{proposition}\label{p#1} } \nc \bco[1] {
\begin{corollary}\label{c#1} } \nc \bde[1] {
\begin{definition}\label{d#1}\rm } \nc \bex[1] {
\begin{example}\label{e#1}\rm } \nc \bre[1] {
\begin{remark}\label{r#1}\rm } \nc \bcon[1] {
\medskip\noindent{\it{Conjecture #1}} } \nc \bqu[1]  {
\medskip\noindent{\it{Question #1}} }
\nc {\ethe} { \end{theorem} }
 \nc {\ele} { \end{lemma} } \nc {\epr}
{ \end{proposition} } \nc {\eco} { \end{corollary} } \nc {\ede} {
\end{definition} } \nc {\eex} { \end{example} } \nc {\ere} {
\end{remark} } \nc {\econ} {\smallskip} \nc {\equ} {\smallskip}
 \nc \thref[1]{Theorem \ref{t#1}}
\nc \leref[1]{Lemma \ref{l#1}} \nc \prref[1]{Proposition
\ref{p#1}} \nc \coref[1]{Corollary \ref{c#1}} \nc
\deref[1]{Definition \ref{d#1}} \nc \exref[1]{Example \ref{e#1}}
\nc \reref[1]{Remark \ref{r#1}}

\def \T {{\mathcal T}}

\def \diag { {\mathrm{diag}} }

 \def\AA  {\kern-0.1em}
 \def\BB  {\kern+0.1em}
 \def\BBB {\kern+0.15em}
 \def\K   {\kern+0.05em}
 \def\MK  {\kern-0.07em}
 \def\MKK {\kern-0.04em}

\begin{document}

\vspace{0.5cm}

\title[ Non-integrability of the Sasano system ]
{ Non-integrability of the rational Sasano system of type  $A_4^{(2)}$ }

\author[Tsvetana  Stoyanova]{Tsvetana  Stoyanova}

\date{07.05.2024}

 \maketitle

\begin{center}
{Department of Mathematics and Informatics,
Sofia University,\\ 5 J. Bourchier Blvd., Sofia 1164, Bulgaria, 
cveti@fmi.uni-sofia.bg}
\end{center}

\vspace{0.5cm}

\begin{abstract}
    In this paper we study the integrability the Sasano system of type $A^{(2)}_4$ from the point of view of the Hamiltonian dynamics. We  prove rigorously that for all values of the parameters
      for which the Sasano system of type $A^{(2)}_4$ has a particular rational solution it is not integrable by rational  first integrals.  By an explicit computation we show that the connected component $G^0$
   of the unit element  of the differential Galois group of the  normal variational equations along a simple particular rational solution is a direct product of two groups $SL_2(\CC)$.
   Thus the Morales-Ramis theory leads to a non-integrable result. Moreover, using B\"acklund transformations we extend the obtained particular
   non-integrable result to the entire orbit of the parameters.
	 \end{abstract}

   \maketitle



\headsep 10mm \oddsidemargin 0in \evensidemargin 0in

\section{Introduction}

 The Sasano systems are higher order  Painlev\'{e} systems which admit certain affine Weyl groups of symmetry.
  They were introduced by Yusuke Sasano in a series of papers \cite{Sa, Sa1, Sa2, Sa3, Sa4, Sa5, Sa6} where for a given Weyl group of symmetry $W(A)$ he  found a higher order
  system of nonlinear ordinary differential equations for  which $W(A)$ acts as its  B\"acklund transformations. The Sasano systems can be considered as higher order
analogues of the classical second-order Panlev\'{e} equations. Although they were introduced by strictly mathematical consideration the Sasano systems turn out to find applications
in mathematical physics through the Drinfeld-Sokolov hierarchies \cite{FS}.

  One of the essential properties of the Sasano systems is that all of them are time dependent Hamiltonian systems. The question about their integrability from the point of view of
the Hamiltonian dynamics then naturally arises. In \cite{St4} we have proved that the Sasano system of type $D^{(1)}_5$ is not integrable with values of the parameters
 $\alpha_1=\alpha_2=\alpha_3=0,\, \alpha_0+\alpha_5=0$.  In this paper we study the integrabilty of the Sasano system of type $A^{(4)}_2$ given by \cite{Sa1}
  \be\label{s}
   \dot{x} &=&
   4\,x\,y - 2 \alpha_1 + 2\,z\,w,\nonumber\\[0.15ex]
    \dot{y} &=&
    -2\,y^2 - 4\,x -2\,t - w,\\[0.15ex]
   \dot{z} &=&
   z^2  - w + x + 2\,y\,z,\nonumber\\[0.15ex]
    \dot{w} &=&
    -2\, z\, w - \alpha_0 - 2 \,y\,w\,,\nonumber
  \ee
 where  $\alpha_j, 0 \leq j \leq 2$
 are complex parameters which satisfy the following relation
   \be\label{r}
    \alpha_0+ 2 \alpha_1 + 2 \alpha_2=1
    \ee
  The Sasano system \eqref{s} admits the affine Weyl group $W(A^{(4)}_2)$ as a group of B\"acklund transformations \cite{Sa1}.
  The  system \eqref{s} is a time-dependent  Hamiltonian system of $2+1/2$ degrees of freedom \cite{Sa1}
   \ben
    \frac{d x}{d t}=\frac{\partial H}{\partial y},\quad
    \frac{d y}{d t}=-\frac{\partial H}{\partial x},\quad
    \frac{d z}{d t}=\frac{\partial H}{\partial w},\quad
    \frac{d w}{d t}=-\frac{\partial H}{\partial z}\,,
   \een
   with the polynomial Hamiltonian
   \ben
     H   &=& 2\,x\,y^2 + 2\,x^2 + 2\,t\,x - 2\,\alpha_1\,y + z^2\,w - \frac{w^2}{2} + \alpha_0\,z
     +x\,w + 2\,y\,z\,w=\\[0.15ex]
             &=&
            2\,H_{II}(x, y, t ; \alpha_1) + H_{II}^{\textrm{auto}}(z, w, ; \alpha_0) + x\,w + 2 \,y\,z\,w,.
     \een
   The symbol $H_{II}$ denotes the Hamiltonian associated to the second Painlev\'{e} equation given in the form
         $$\,
          H_{II}(x, y, t ; \alpha_1)=x\,y^2 +x^2 + t\,x-\alpha_1\,y\,,
         \,$$
          and the symbol $H_{II}^{\textrm{auto}}$ denotes the autonomous version of $H_{II}$ given by
                   $$\,
                     H_{II}^{\textrm{auto}}(z, w ; \alpha_0)=z^2\,w - \frac{w^2}{2} + \alpha_0\,z\,.
                      \,$$
             In this way the Sasano system \eqref{s} can be considered as a coupled Painlev\'{e} II equation.
      The system \eqref{s} can be turned into an autonomous Hamiltonian system with three degrees of freedom by introducing two new dynamical variables : $t$ and a conjugate
   to it $F$. The new Hamiltonian becomes
                    $$\,
                         \widehat{H}=H+F
                      \,$$
    and the associated Hamiltonian system is
        \be\label{H}
                               & &
    \frac{d x}{d s}=\frac{\partial \widehat{H}}{\partial y},\quad
    \frac{d y}{d s}=-\frac{\partial \widehat{H}}{\partial x},\quad
    \frac{d z}{d s}=\frac{\partial \widehat{H}}{\partial w},\quad
    \frac{d w}{d s}=-\frac{\partial \widehat{H}}{\partial z},\\[0,25ex]
                             & &
      \frac{d t}{d s}=\frac{\partial \widehat{H}}{\partial F},\quad
    \frac{d F}{d s}=-\frac{\partial \widehat{H}}{\partial t}\,.\nonumber
   \ee
              The symplectic structure $\Omega$ is canonical in the variables $(x, y, z, w, t, F)$, that is
         $\Omega=d y \wedge d x + d w \wedge d z + d F \wedge d t$.

In this paper we strictly prove that for these values of the parameters $\alpha_j, j=0, 1, 2$ for which the Sasano system
   \eqref{s} admits a particular rational solution the Hamiltonian system \eqref{H} is not integrable by rational first integrals. Recall that according
to the theorem of Liouville-Arnold  \cite{Ar} this means non-existence of three independent rational first integrals in involution.
  As in our works on the non-integrability of the second order Painlev\'{e} equation \cite{HS, St1, St2, St3} and of the Sasano system of type $D^{(1)}_5$ \cite{St4} in this paper we utilize the
Morales-Ramis theory of non-integrability of Hamiltonian systems. This theory transforms the problem of integrability of a given complex analytical Hamiltonian system in the
sense of the theorem of Liouville-Arnold to the problem of integrability of a linear system of ordinary differential equations in the sense of the differential Galois theory.

   From the paper \cite{Mat} of Matsuda it follows that all the rational solutions of the Sasano system  \eqref{s} can be obtained by B\"acklund transformations from the
following simple particular rational solution
   \be\label{sol}
    x=w=-\frac{2}{5}\,t,\quad y=z=0, \quad \alpha_0=\frac{2}{5},\quad \alpha_1=\frac{1}{5}, \quad \alpha_2=\frac{1}{10}\,.
  \ee
In this paper studying the normal variational equations of \eqref{H} along the solution \eqref{sol} we establish a non-integrable result. More precisely,
by successive shearing and similar transformations we reduce the normal variational equations, which a fourth order linear system, to a system consisted of two independent
second order linear systems. Each of these systems is equivalent to a non-solvable Whittaker equation. This fact implies that the connected component $G^0$ of the unit element
of the differential Galois group $G$ of the normal variational equations has the form $G^0=SL_2(\CC) \times SL_2(\CC)$ which is not Abelian.
  Then the key result of this paper is the following theorem
    \bth{1}
     Assume that
                 $$\,
         \alpha_0=\frac{2}{5}, \quad \alpha_1=\frac{1}{5},\quad \alpha_2=\frac{1}{10}\,.
              \,$$
      Then the Sasano system of type $A^{(2)}_4$  \eqref{s} is not integrable by rational first integrals.
   \ethe
   Since  the B\"acklund transformations of the Sasano system of type $A^{(2)}_4$ are canonical transformations which are rational in all of the dynamical variables \cite{Sa1}
 we can extend the key result of \thref{1} to the main theorem of this paper
     \bth{main}
             For all values of the parameters for which  the Sasano system of type $A^{(2)}_4$ admits a particular rational solution,  it is not integrable by rational first integrals.
     \ethe

   This paper is organized as follows. In the next section we briefly recall the main theorem of the Morales-Ramis theory of non-integrability of Hamiltonian systems.
   In Section 3 we prove the key \thref{1}. In the last section using the B\"acklund transformations of the Sasano system \eqref{s} we establish the main result of this paper.
         \section{Preliminaries}

      In this section we briefly recall the Morales-Ramis theory on non-integrability of Hamiltonian systems following \cite{M, MR}.

   Consider a complex analytical Hamiltonian system
          \be\label{HS}
            \dot{x}=X_H(x)
         \ee
        with a Hamiltonian $H : M \rightarrow \CC$ on a complex symplectic manifold $M$ of dimension $2 n$.  Recall that by the theorem of Liouville-Arnold \cite{Ar}
    the Hamiltonian system \eqref{HS} is completely integrable if there exist  $n$  functionally independent first integrals $f_1=H, f_2, f_3, \ldots$ and in involution.
    Let $x(t)$ be a non-equilibrium particular solution of \eqref{HS}.
       denote by $\Gamma$ the phase curve corresponding to this solution. The variational equations $(\textrm{VE})$ along $\Gamma$ have the form
         $$\,
              \dot{\xi}=\frac{\partial X_{H}(x(t))}{\partial x}\,\xi, \quad \xi\in T_{\Gamma} M\,.
       \,$$
         We always can reduce the variational equations using the Hamiltonian in the following sense. Consider the normal bundle of $\Gamma$ on the level variety
    $M_h=\{x\,|\,H(x)=h\}$. The projection of the variational equations on this bundle is called the normal variational equations $(\textrm{NVE})$. The dimension of the $(\textrm{NVE})$
 is $2 n -2$.  The solutions of the $(\textrm{NVE})$ define a Picard-Vessiot extension $L$ of the basic differential field $K$ of coefficients of the $(\textrm{NVE})$. This in its turn defines
   a differential Galois group $G=\textrm{Gal}(L/K)$. Then the main theorem of the Morales-Ramis theory states \cite{M, MR}
        \bth{MR}(Morales-Ruiz and Ramis)
           If the Hamiltonian system \eqref{HS} is completely integrable by meromorphic first integrals in a neighborhood of $\Gamma$, not necessarily independent on $\Gamma$
              itself, then the connected component $G^0$ of the unit element of the differential Galois group $G=\textrm{Gal}(L/K)$ is Abelian.
         \ethe
  \thref{MR} provides a necessary condition for completely integrability and a sufficient condition for non-integrability of a given Hamiltonian system. That is, if the connected component of the unit
 element of the Galois group $G$ is not Abelian, then the corresponding Hamiltonian system \eqref{HS} is not integrable by meromorphic first integrals.


   \section{Non-integrability for $(\alpha_0, \alpha_1, \alpha_2)=(2/5, 1/5, 1/10)$}

    In this section we will prove that when $(\alpha_0, \alpha_1, \alpha_2)=(2/5, 1/5, 1/10)$ the extended autonomous Sasano system \eqref{H}
    is not integrable by meromorphic first integrals which are rational functions in $t$.
    To obtain this result we will study the variational equations along the basic rational solutions
      \be\label{rs}
      x=-\frac{2}{5}\,s, \quad y=0,\quad z=0, \quad w=-\frac{2}{5}\,s, \quad t=s, \quad F=\frac{2}{5}\,s^2\,.
      \ee
          Sinse $t=s$ from here on we use $t$ as an independent variable instead of $s$.

   For the  normal variational equations $(\textrm{NVE})$ of the  system \eqref{H} along the particular solution
   \eqref{rs} we obtain the system
    \ben
      \dot{x}_1  &=&
       - \frac{8}{5}\,t\,y_1 - \frac{4}{5}\,t\,z_1,\\[0.15ex]
       \dot{y}_1 &=&
       -4\,x_1 - w_1,\\[0.15ex]
       \dot{z}_1 &=&
       x_1 - w_1,\\[0.15ex]
        \dot{w}_1 &=&
        \frac{4}{5}\,t\,y_1 + \frac{4}{5}\,t\,z_1\,.
    \een
   The $(\textrm{NVE})$ have one singular point over $\CC\PP^1$ at $t=\infty$ which is an irregular singularity.
  The origin is one ordinary point for it.

    The so obtained $(\textrm{NVE})$ can be written in the form
     \be\label{eq1}
      t^{-1}\,\dot{X}(t)=A(t)\,X(t)\,,
     \ee
    where $X(t)$ is the vector-column $X(t)=(x_1(t), y_1(t), z_1(t), w_1(t))$ and $A(t)$ is the holomophic matrix at
    $t=\infty$
      \be\label{a}
       A(t)=\left(\begin{array}{rrrr}
         0    & -\frac{8}{5}   & -\frac{4}{5}  & 0\\[0.15ex]
         -\frac{4}{t}   & 0    & 0   & -\frac{1}{t}\\[0.15ex]
          \frac{1}{t}   & 0    & 0   & - \frac{1}{t}\\[0.15ex]
           0   &\frac{4}{5}    & \frac{4}{5}   & 0
              \end{array}\right)\,.
      \ee

      Denote by $M_n(t)$ the set of all square matrices of order $n$ whose elements are functions in $t$.

      \bth{reduction}
       There exist a matrix $S(t)\in M_4(t)$ whose elements are algebraic functions in $t$ and an algebraic function $f(\tau)$ in $\tau$,
       such that the
        transformation
        \be\label{t}
           X(t)=S(t)\,V(t), \qquad t=f(\tau)
        \ee
       reduces the system \eqref{eq1} to the system
        \be\label{eq}
         \tau^{-5}\,V'(\tau)=Q(\tau)\,V(\tau),\quad '=\frac{d}{d \tau}\,.
        \ee
        The matrix $Q(\tau)$ is a holomorphic matrix at $\tau=\infty$ which has the form
         $$\,
          Q(\tau)=\left(\begin{array}{cc}
             Q_1(\tau)  & 0\\
              0         & Q_2(\tau)
              \end{array}\right)
         \,$$
         with $Q_j(\tau)\in M_2(\tau), j=1, 2$.
      \ethe

     \proof

       In order to simplify the system \eqref{eq1} we apply to it shearing transformations at the irregular singularity $t=\infty$. For details about such transformations
      the reader can look at the book of Wasow \cite{W}.

       All the eigenvalues of matrix $A(\infty)$ are equal to zero. The transformation $X(t)=T_1\,V_1(t)$ with
        $$\,
          T_1=\left(\begin{array}{crcr}
           4   & 0   & 0   & 0\\
           0   & -5  & 0   & -5\\
           0   & 5   & 0   & 10\\
           0   & 0   & 4   & 0
           \end{array}\right),\qquad
           T_1^{-1}=\left(\begin{array}{crrc}
           \frac{1}{4}   & 0   & 0   & 0\\[0.15ex]
           0   & -\frac{2}{5}  & -\frac{1}{5}   & 0\\[0.15ex]
           0   & 0   & 0   & \frac{1}{4}\\[0.15ex]
           0   & \frac{1}{5}   & \frac{1}{5}   & 0
           \end{array}\right)
        \,$$
         takes the system \eqref{eq1} into the system
          \be\label{eq2}
            t^{-1}\,\dot{V}_1(t)=T_1^{-1}\,A(t)\,T(t)=B(t)\,V_1(t)=\left[B_0 + B_1\,t^{-1}\right]\,V_1(t)
          \ee
          with
          $$\,
           B_0=\left(\begin{array}{cccc}
            0   & 1   & 0   & 0\\
            0   & 0   & 0   & 0\\
            0   & 0   & 0   & 1\\
            0   & 0   & 0   & 0
            \end{array}\right),\qquad
            B_1=\left(\begin{array}{rcrc}
               0   & 0   & 0   & 0\\[0.15ex]
               \frac{28}{5}  & 0   & \frac{12}{5}   & 0\\[0.15ex]
               0   & 0   & 0   & 0\\[0.15ex]
               -\frac{12}{5}  & 0  & -\frac{8}{5}   & 0
               \end{array}\right)\,.
          \,$$
     The first shearing transformation $V_1(t)=S_1(t)\,V_2(t)$ with $S_1(t)=\diag (1, t^{-g}, t^{-2 g}, t^{-3 g})$ takes the system
     \eqref{eq2} into the system
      \be\label{eq3}
       t^{-1}\,\dot{V}_2(t)=\left[S^{-1}(t)\,B(t)\,S(t) - t^{-1}\,S^{-1}(t)\,\dot{S}(t)\right]\,V_2(t)=P(t)\,V_2(t)\,,
      \ee
      where
       $$\,
        P(t)=\left(\begin{array}{cccc}
          0   & t^{-g}  & 0   & 0\\[0.15ex]
          \frac{28}{5}\,t^{g-1}   & g\,t^{-2}    & \frac{12}{5}\,t^{-g-1}    & 0\\[0.15ex]
          0    & 0    & 2 g\,t^{-2}   & t^{-g}\\[0.15ex]
          - \frac{12}{5}\,t^{3 g-1}   & 0   & -\frac{8}{5}\,t^{g-1}   & 3 g\,t^{-2}
          \end{array}\right)\,.
       \,$$
         With $g=1/4$ the system \eqref{eq3} becomes
          $$\,
           t^{-1}\,\dot{V}_2(t)=
           \left(\begin{array}{cccc}
            0   & t^{-1/4}    & 0    & 0 \\[0.15ex]
            \frac{28}{5}\,t^{-3/4}   & \frac{1}{4}\,t^{-2}   & \frac{12}{5}\,t^{-5/4}    & 0\\[0.15ex]
            0    & 0    &\frac{1}{2}\,t^{-2}    & t^{-1/4}\\[0.15ex]
            -\frac{12}{5}\,t^{-1/4}  & 0    & -\frac{8}{5}\,t^{-3/4}   & \frac{3}{4}\,t^{-2}
            \end{array}\right)\,V_2(t)\,.
          \,$$
          Now setting $t=\alpha\,\tau^4$ where $\alpha$ will be chosen late and multiplying by $4\,\alpha^2\,\tau$ we get
           \be\label{eq4}
            \tau^{-6}\,V_2'(\tau)=P(\tau)\,V_2(\tau),\qquad '=\frac{d}{d \tau}\,,
           \ee
           where
            $$\,
              P(\tau)=\left(\begin{array}{cccc}
               0   & 4\,\alpha^{7/4}   & 0   & 0\\[0.15ex]
               \frac{4.28}{5}\,\alpha^{5/4}\,\tau^{-2}     & \tau^{-7}   & \frac{48}{5}\,\alpha^{3/4}\,\tau^{-4}    & 0 \\[0.15ex]
               0    & 0   & 2\, \tau^{-7}   & 4\,\alpha^{7/4} \\[0.15ex]
               -\frac{48}{5}\,\alpha^{7/4}    & 0    & -\frac{32}{5}\,\alpha^{5/4}\,\tau^{-2}    & 3\,\tau^{-7}
               \end{array}\right)\,.
            \,$$
           Again all the eigenvalues of the leading matrix $P(\infty)$ are equal to zero. But this time the matrix $P(\infty)$ is  similar
           to a Jordan block of order $4$. The similar  transformation $V_2(\tau)=T_2\,V_3(\tau)$ with
           $$\,
            T_2=\left(\begin{array}{cccc}
               0   & 0  & -\frac{5}{48}\,\alpha^{-7/4}   & 0\\[0.15ex]
               0   & 0  & 0                              & -\frac{5}{4.48}\,\alpha^{-7/2}\\[0.15ex]
               4\,\alpha^{7/4}  & 0   & 0  & 0\\ [0.15ex]
               0   & 1  & 0    & 0
               \end{array}\right)
             \,$$
       and
       $$\,
              T_2^{-1}=\left(\begin{array}{cccc}
              0   & 0  & \frac{1}{4}\,\alpha^{-7/4}   & 0\\[0.15ex]
              0   & 0  & 0                              & 1\\[0.15ex]
              -\frac{48}{5}\,\alpha^{7/4}  & 0   & 0  & 0\\ [0.15ex]
              0   & -\frac{4.48}{5}\,\alpha^{7/2}  & 0    & 0
              \end{array}\right)
           \,$$
           takes the system \eqref{eq4}  into the system
            \be\label{eq5}
             \tau^{-6}\,V'_3(\tau)=(T^{-1}_2\,P(t)\,T_2)\,V_3(\tau)=P_1(\tau)\,V_3(\tau)\,,
            \ee
            where
            $$\,
             P_1(\tau)=\left(\begin{array}{cccc}
             2\,\tau^{-7}    & 1     & 0    & 0\\[0.15ex]
             -\frac{4.32}{5}\,\alpha^3\,\tau^{-2}   & 3 \,\tau^{-7}   & 1   & 0\\[0.15ex]
              0     & 0    & 0   & 1\\[0.15ex]
              -\frac{16.48.48}{25}\,\alpha^6\,\tau^{-4}   & 0   & \frac{64.7}{5}\,\alpha^3\,\tau^{-2}   & \tau^{-7}
              \end{array}\right)\,.
            \,$$
            A shearing transformation
            $$\,
             V_3(\tau)=\left(\begin{array}{cccc}
              1    & 0    & 0    & 0\\
              0    & \tau^{-1}   & 0    & 0\\
              0    & 0    & \tau^{-2}   & 0\\
              0    & 0    & 0           & \tau^{-3}
              \end{array}\right)\,V_4(\tau)
            \,$$
            with $g=1$ produces the system
            \be\label{eq6}
             \tau^{-5}\,V'_4(\tau)=P_2(\tau)\,V_4(\tau)\,,
            \ee
            where
            $$\,
              P_2(\tau)=\left(\begin{array}{cccc}
               2\,\tau^{-6}    & 1    & 0    & 0\\[0.15ex]
               - \frac{4.32}{5}\,\alpha^3     & 4\,\tau^{-6}    & 1    & 0\\[0.15ex]
               0              & 0     & 2\,\tau^{-6}    & 1\\[0.15ex]
               -\frac{16.48.48}{25}\,\alpha^6  & 0   &\frac{64.7}{5}\,\alpha^3    & 4\,\tau^{-6}
               \end{array}\right)\,.
            \,$$
        Choosing $\alpha^3=\frac{5}{64}$ we find that the leading matrix
        $$\,
         P_2(\infty)=\left(\begin{array}{rccc}
          0    & 1    & 0   & 0\\
          -2   & 0    & 1   & 0\\
          0    & 0    & 0   & 1\\
          -9   & 0    & 7   & 0
          \end{array}\right)
         \,$$
         has four distinct eigenvalues
        \be\label{l}
          & &
          \lambda_1=-\frac{i\,\sqrt{6\,\sqrt{5} -10}}{2},\quad
          \lambda_2=\frac{i\,\sqrt{6\,\sqrt{5} -10}}{2},\\[0.2ex]
          & &
          \lambda_3=-\frac{\sqrt{6\,\sqrt{5} +10}}{2},\quad
          \lambda_4=\frac{\sqrt{6\,\sqrt{5} +10}}{2}\,.\nonumber
         \ee
       Now the transformation $V_4(\tau)=T_3\,V(\tau)$
       with
        $$\,
          T_3=\left(\begin{array}{cccc}
          \frac{\sqrt{5}\,(\sqrt{5}+3)}{20}    &  \frac{\sqrt{5}\,(\sqrt{5}+3)}{20}    &  \frac{\sqrt{5}\,(\sqrt{5}-3)}{20}
           & \frac{\sqrt{5}\,(\sqrt{5}-3)}{20}\\[0.4ex]
         - \frac{i}{\sqrt{6 \sqrt{5}-10}}      &  \frac{i}{\sqrt{6 \sqrt{5}-10}}  & \frac{1}{\sqrt{6 \sqrt{5}+10}}
          & - \frac{1}{\sqrt{6 \sqrt{5}+10}}\\[0.4ex]
          \frac{3 \sqrt{5}}{10}    & \frac{3 \sqrt{5}}{10}   & -\frac{3 \sqrt{5}}{10}   & - \frac{3 \sqrt{5}}{10}\\[0.4ex]
          -\frac{3\,i\,\sqrt{5}\,\sqrt{6 \sqrt{5}-10}}{20}     & \frac{ 3\,i\,\sqrt{5}\,\sqrt{6 \sqrt{5}-10}}{20}
          & \frac{ 3\,\sqrt{5}\,\sqrt{6 \sqrt{5}+10}}{20}  &  -\frac{ 3\,\sqrt{5}\,\sqrt{6 \sqrt{5}+10}}{20}
               \end{array}\right),
        \,$$
        and
        $$\,
        T_3^{-1}=\left(\begin{array}{cccc}
         1    & \frac{2\,i}{\sqrt{6 \sqrt{5}-10}}  & \frac{\sqrt{5}-3}{6}
         & -\frac{i\,\sqrt{5}\,\sqrt{6 \sqrt{5}-10}}{30}\\[0.4ex]
           1    & -\frac{2\,i}{\sqrt{6 \sqrt{5}-10}}  & \frac{\sqrt{5}-3}{6}
           & \frac{i\,\sqrt{5}\,\sqrt{6 \sqrt{5}-10}}{30}\\[0.4ex]
           1    & -\frac{2}{\sqrt{6 \sqrt{5}+10}}  & -\frac{\sqrt{5}+3}{6}
           & \frac{\sqrt{5}\,\sqrt{6 \sqrt{5}+10}}{30}\\[0.4ex]
             1    & \frac{2}{\sqrt{6 \sqrt{5}+10}}  & -\frac{\sqrt{5}+3}{6}
             & -\frac{\sqrt{5}\,\sqrt{6 \sqrt{5}+10}}{30}
             \end{array}\right)\,.
        \,$$
       changes the differential equation \eqref{eq6} into
       the equation
        \be\label{eq7}
         \tau^{-5}\,V'(\tau)=\left(\begin{array}{cccc}
         	\lambda_1 + 3\,\tau^{-6}   & -\tau^{-6}     & 0    & 0\\[0.15ex]
         	 -\tau^{-6}   &\lambda_2 + 3\,\tau^{-6}     & 0    & 0\\[0.15ex]
         	 0    & 0   & \lambda_3 + 3\,\tau^{-6}    & -\tau^{-6}\\[0.15ex]
         	 0    & 0   & -\tau^{-6}        &\lambda_4 + 3\,\tau^{-6}
         	\end{array}\right)\,V(\tau)\,,
        \ee
  with $\lambda_j, j=1, 2, 3, 4$ defined by \eqref{l}.
         The system \eqref{eq7} is the wanted system \eqref{eq}.

         This ends the proof.
         \qed

      \bre{al}
       Following the Wasow's book we have to put $\alpha=4^{-4/7}$. Here we choose $\alpha^3=5/64$ in order to simplify the eigenvalues of the matrix $P_2(\infty)$.
    Our choice is allowed since it does not change the aim of the shearing transformations, which is to obtain a system of differential equations with a leading matrix
different from $A(\infty)$.
    \ere

   The resulting system \eqref{eq7} has two singular points over $\CC\PP^1$ : $\tau=0$ and $\tau=\infty$,  while the $(\textrm{NVE})$ have only one at $t=\infty$. In  the next proposition
   we show that $\tau=0$ is an apparent singularity for the system \eqref{eq7}. This phenomenon seems to be similar to that one observed at the reduction of second order linear system
  of ordinary differential equations to a second order scalar equation. Recall that in this case if the element $a_{12}(t)$ of the matrix $A(t)=\{a_{ij}(t)\}_{i, j=1}^2,\,a_{ij}\in\CC(t)$ for the
   system $\dot{x}=A(t)\,x$ has a zero of second order at $t=a$ then $t=a$ is an apparent singularity for the resulting second order scalar equation.

           \bpr{0}
           The point $\tau=0$ is an apparent singularity for the system \eqref{eq7}.
           \epr

           \proof

              The change of the independent variable
              $$\,
              \tau^6=\eta
              \,$$
              takes the differential equation \eqref{eq7} into the equation
              \be\label{eq8}
              \frac{d V(\eta)}{d \eta}=\left(\begin{array}{cccc}
              	\frac{\lambda_1}{6}+\frac{1}{2 \eta}    & - \frac{1}{6 \eta}    & 0   & 0\\[0.2ex]
              	- \frac{1}{6 \eta}     & \frac{\lambda_2}{6} + \frac{1}{2 \eta}    & 0   & 0\\[0.2ex]
              	0   & 0    & \frac{\lambda_3}{6} + \frac{1}{2 \eta}    & - \frac{1}{6 \eta}\\[0.2ex]
              	0   & 0    & -\frac{1}{6 \eta}   & \frac{\lambda_4}{6} + \frac{1}{2 \eta}
              \end{array}\right)\,V(\eta)\,.
              \ee
              The system \eqref{eq8} is reduced to two independent second-order linear systems in the form
              \be\label{12}
              \frac{d u}{d \eta}=\left(\begin{array}{cc}
              	\frac{\lambda_i}{6} + \frac{1}{2 \eta}     & - \frac{1}{6 \eta}\\[0.25ex]
              	- \frac{1}{6 \eta}                   & \frac{\lambda_j}{6} + \frac{1}{2 \eta}
              \end{array}\right)\,u
              \ee
              for $(i, j)=(1, 2)$ or $(i, j)=(3, 4)$.
                Let $u=(u_1, u_2)^T$. Since $\lambda_1+\lambda_2=\lambda_3+\lambda_4=0$ the system \eqref{12}
                is equivalent to the second-order scalar equation
                \be\label{e1}
                \frac{d^2 u_1}{d \eta^2} -
                \left[-\frac{\lambda_i\,\lambda_j}{36} + \frac{\lambda_i-\lambda_j}{12\,\eta} - \frac{2}{9\,\eta^2}\right]\,u_1=0\,.
                \ee
           The point $\eta=0$ is a regular singularity for the equation \eqref{e1}. The characteristic equation
           for $\eta=0$ is given by
           $$\,
           \rho\,(\rho-1) + \frac{2}{9}=0\,.
           \,$$
           The corresponding characteristic exponents are $\rho_1=\frac{2}{3}$ and $\rho_2=\frac{1}{3}$. Since
           $\rho_1-\rho_2\notin\ZZ$ the local theory of regular singularities ensures that the equation \eqref{e1}
           has two independent local solutions $u_1(\eta)$ and $\widetilde{u}_1(\eta)$ near the origin in the form
           $$\,
           u_1(\eta)=\eta^{2/3}\,f_1(\eta) \quad \textrm{and} \quad
           \widetilde{u}_1(\eta)=\eta^{1/3}\,f_2(\eta)\,.
           \,$$
           The functions $f_1(\eta)$ and $f_2(\eta)$ are holomorphic functions near the origin.
           Then the system \eqref{eq8} admits a local fundamental matrix solution $\Psi(\eta)$ near the origin $\eta=0$
           in the form
           $$\,
           \Psi(\eta)=U(\eta)\,\eta^J\,,
           \,$$
           where $U(\eta)$ is a holomorphic matrix-function near the origin, and
           $$\,
           J=\diag\left(\frac{2}{3}, \frac{1}{3}, \frac{2}{3}, \frac{1}{3}\right)\,.
           \,$$
           Therefore the system \eqref{eq7} admits a local fundamental matrix solution near the origin in the form
           $$\,
           \Psi(\tau)=U(\tau^6)\,\tau^{J_1}\,,
           \,$$
           where $J_1=6\,J=\diag (4, 2, 4, 2)$. Thus the origin $\tau=0$ is an apparent singularity for the system
           \eqref{eq7}.

           This ends the proof.
           \qed

      \bpr{G}
       The connected component $(G_1)^0$ of the unit element of the differential Galois group $G_1$ of the system \eqref{eq7} is a direct product of two groups $SL_2(\CC)$,
       i.e. it has the form $(G_1)^0=SL_2(\CC) \times SL_2(\CC)$.
      \epr

      \proof

        We will show that  the local differential Galois group at $\eta=\infty$  of  each systems \eqref{12} coincides with $\textrm{SL}_2(\CC)$.
      The change of the variable
       \be\label{c1}
         \eta=\frac{3}{\sqrt{-\lambda_i\,\lambda_j}}\,\zeta
       \ee
     takes the equation \eqref{e1} into the equation
      \be\label{e2}
        \frac{d^2 u_1}{d \zeta^2} -
        \left[\frac{1}{4} - \frac{\lambda_j-\lambda_i}{4\,\sqrt{-\lambda_i\,\lambda_j}}\,\frac{1}{\zeta}
         - \frac{2}{9 \,\zeta^2}\right]\,u_1=0.
      \ee
      The equation \eqref{e2} is the Whittaker equation \cite{WW}
       $$\,
        \frac{d^2 u_1}{d \zeta^2}
        - \left[\frac{1}{4}  - \frac{\kappa}{\zeta} + \frac{4\,\mu^2-1}{4 \zeta^2}\right]\,u_1=0
       \,$$
       with
       $$\,
        \kappa=\frac{\lambda_j-\lambda_i}{4\,\sqrt{-\lambda_i\,\lambda_j}}=\frac{1}{2},\qquad
        \mu=\frac{1}{6}\,.
       \,$$
       In \cite{MaR} Martinet and Ramis show that the Whittaker equation has two Stokes matrices at $\zeta=\infty$
        $$\,
         St_{\pi}=\left(\begin{array}{cc}
             1    & 0\\
             \mu_1   & 1
             \end{array}\right),\quad
            St_{0}=\left(\begin{array}{cc}
            1    & \mu_2\\
            0   & 1
            \end{array}\right)\,,
        \,$$
        which correspond to the singular directions $\theta_1=\pi$ and $\theta_2=0$. In Proposition 3.3.1 in \cite{MaR} they
        prove that
         \begin{enumerate}
         	\item\,
         	$\mu_1=0$ if and only if $\kappa-\mu\in\frac{1}{2} + \NN$ or $\kappa+\mu\in\frac{1}{2} + \NN$
         	
         	\item\,
         	$\mu_2=0$ if and only if $-\kappa-\mu\in\frac{1}{2} + \NN$ or $-\kappa+\mu\in\frac{1}{2} +\NN$.
         \end{enumerate}
         Since  $\kappa \pm \mu=\frac{1}{2} \pm \frac{1}{6}\notin \frac{1}{2}+\NN$ and
         $-\kappa\pm \mu=-\frac{1}{2} \pm \frac{1}{6}\notin\frac{1}{2} + \NN$ we have that the equation
         \eqref{e2} has two non-trivial Stokes matrices. Since the exponential torus $\T$ are given by \cite{MaR}
         $$\,
          \T=\left\{\left(\begin{array}{cc}
              a   & 0\\
              0   & a^{-1}
              \end{array}\right)\,|\,a\in\CC^*\right\}
         \,$$
         we conclude that the local differential Galois group of the equation \eqref{e2} at $\zeta=\infty$ is the group $SL_2(\CC)$, \cite{MaR} .
        The substitution \eqref{c1} changes the local Galois group at $\zeta=\infty$ of the equation \eqref{e1} but remains its connected component of the unit element.
      Thus the connected component of the unit element of the local differential Galois group at $\eta=\infty$ of the system \eqref{12} is the group $SL_2(\CC)$.
        Therefore connected component of the unit element of the local differential Galois group at $\eta=\infty$ of the systems \eqref{eq8}  is a direct product of two groups $SL_2(\CC)$,
         i.e. it has the form
         $SL_2(\CC) \times SL_2(\CC)$.
        From \prref{0} it follows that the differential Galois group of the system \eqref{eq7} is its local Galois group at $\tau=\infty$.
    The transformation $\tau^6=\eta$  changes the  the differential Galois group $G_1$ of the system \eqref{eq7} but its connected component $(G_1)^0$ of the unit element is preserved .
     Thus  $(G_1)^0=SL_2(\CC) \times SL_2(\CC)$.

         This ends the proof.
         \qed

         \bre{r1}
           In fact both of the systems \eqref{12} are not equivalent to the Whittaker equation \eqref{e2} in the same variable $\zeta$.
         The transformation
             $$\,
               \eta=\frac{3}{\sqrt{-\lambda_1\,\lambda_2}}\,\zeta
             \,$$
             takes only the system \eqref{12} for $(i, j)=(1, 2)$ into the equation \eqref{e2}. At the same time this transformation
             takes the second system for $(i, j)=(3, 4)$ into the equation
              \be\label{ne}
                \frac{d^2 u_1}{d \zeta^2} -
                \left[ \frac{7 + 3\,\sqrt{5}}{8} + \frac{i\,(3+\sqrt{5})}{4}\,\frac{1}{\zeta} - \frac{2}{9\,\zeta^2}\right]\,u_1=0.
              \ee
              However, the equation \eqref{ne} is not solvable and its differential Galois group is $SL_2(\CC)$ since the
              change
                   $$\,
                       \zeta=\sqrt{\frac{7 - 3\,\sqrt{5}}{2}}\,\omega
                       \,$$
          takes it into the non-solvable     Whittaker equation
                    $$\,
                                     \frac{d^2 u_1}{d \omega^2} -
                \left[ \frac{1}{4} + \frac{i}{2 \omega} - \frac{2}{9\,\omega^2}\right]\,u_1=0.
                     \,$$

         \ere

      \vspace{1ex}

      Thanks to \prref{G} we establish the main result of this section.

       \bth{key}
         Assume that $\alpha_0=2/5, \alpha_1=1/5, \alpha_2=1/10$. Then the Sasano system of type $A_2^{(4)}$ \eqref{s}
         is not integrable by rational first integrals.
       \ethe

       \proof
       The transformation \eqref{t} defined in the proof of \thref{reduction} changes the differential Galois group $G$ of
       the $(\textrm{NVE})$ but remains the corresponding connected component of the unit element $G^0$ \cite{Si}.
       From \prref{G} it follows that $G^0=SL_2(\CC) \times SL_2(\CC)$ which is not Abelian group. Thus from
       the Morales-Ramis theory it follows that when $\alpha_0=2/5, \alpha_1=1/5, \alpha_2=1/10$ the Sasano system \eqref{s} is not integrable by rational first integral.

       This ends the proof.
       \qed


     \section{B\"acklund transformations and generalization}
         The Sasano systems  \eqref{s} admits an affine Weyl group of symmetry of type $A^{(2)}_4$ which acts on it as a group of B\"acklund transformations. In this section using the
         B\"acklund transformations of the system \eqref{s} we extend the non-integrable result of \thref{key} to the entire orbit of the parameters $\alpha_0, \alpha_1$ and $\alpha_2$.

          The B\"acklund transformations of the Sasano system \eqref{s} are defined by
          \be\label{bt}
             s_0 \,:\,(x, y, z, w, t ; \alpha_0,  \alpha_1, \alpha_2)   &\rightarrow&
             \left(x, y, z + \frac{\alpha_0}{w}, w, t ; -\alpha_0, \alpha_1+\alpha_0, \alpha_2\right),\\[0.2ex]
            s_1 \,:\,(x, y, z, w, t ; \alpha_0,  \alpha_1, \alpha_2)   &\rightarrow&
             \left(x, y - \frac{\alpha_1}{x+z^2}, z, w - \frac{2 \alpha_1\,z}{x+z^2}, t ; \alpha_0+ 2 \alpha_1, -\alpha_1, \alpha_2+\alpha_1\right),\nonumber\\[0.2ex]
               s_2 \,:\,(x, y, z, w, t ; \alpha_0,  \alpha_1, \alpha_2)   &\rightarrow&
             \left(x + \frac{2 \alpha_2\,y}{f_2}- \frac{\alpha_2^2}{f^2_2}, y- \frac{\alpha_2}{f_2}, z + \frac{\alpha_2}{f_2}, w, t ; \alpha_0, \alpha_1+2 \alpha_2, -\alpha_2\right)\,,\nonumber
          \ee
       where $f_2:=x+y^2 +w+t$.
                 \bre{bt}
            When one of the invariant divisors $f_0=w, f_1=x+z^2$ or $f_2=x+y^2+w+t$ vanishes the parameters become $\alpha_0=0, \alpha_1=0$ or
            $\alpha_2=0$, respectively. So, when one of the divisors $f_i, i=0, 1, 2$ vanishes the corresponding transformation $s_i, i=0, 1, 2$ acts as an identity transformation.
               \ere

         The transformations \eqref{bt} satisfy the relations
           $$\,
                 s_i^2=1,\quad (s_0\,s_1)^4=(s_1\,s_0)^4=1,\quad (s_0\,s_2)^2=(s_2\,s_0)^2=1,\quad (s_1\,s_2)^4=(s_2\,s_1)^4=1.
           \,$$
          Hence the transformations $s_i, i=0, 1, 2$ generate the affine Weyl group $W(A^{(2)}_4)$.

            In \cite{Mat} Matsuda proves that acting on the solution \eqref{rs} by B\"acklund transformations \eqref{bt} we obtain a rational solution of the Sasano system
        \eqref{s}  with parameters satisfying  one of the following condition:
                \be\label{con}
                  5 \alpha_2-1/2 \equiv 0 \,  \textrm{ mod}  \,5,    &\quad& 5 \alpha_1-1 \equiv 0, 2 \, \textrm{mod} \,5,\nonumber\\[0.1ex]
                  5 \alpha_2-1/2 \equiv 1 \,  \textrm{ mod}  \,5,    &\quad& 5 \alpha_1-1 \equiv 2, 3 \, \textrm{mod} \,5,\\[0.1ex]
                  5 \alpha_2-1/2 \equiv 3 \,  \textrm{ mod}  \,5,    &\quad& 5 \alpha_1-1 \equiv 0, 1 \, \textrm{mod} \,5,\nonumber\\[0.1ex]
                  5 \alpha_2-1/2 \equiv 4 \,  \textrm{ mod}  \,5,    &\quad& 5 \alpha_1-1 \equiv 1, 3 \, \textrm{mod} \,5\,.\nonumber
            \ee
                Then we have the main result of this paper
\bth{main1}
               Assume that the parameters $\alpha_i, i=0, 1, 2$ satisfy one of the conditions \eqref{con}. Then the Sasano system of type $A^{(2)}_4$ \eqref{s} is not integrable by rational  first integrals.
        \ethe

             \proof
             The proof follows from the fact that the B\"acklund transformations \eqref{bt} are canonical transformations which are rational in all of the canonical variables \cite{Sa1}.
         \qed

    \section {Conclusion}

    In this paper we strictly prove that when the Sasano system of type $A^{(2)}_4$ admits a particular rational solution it is not integrable by rational  first integrals. 
  The main approach for finding an obstruction to integrability is the Morales-Ramis theory which reduces the problem of integrability of a given Hamiltonian system
     to the problem of the computation of the  differential Galois group of the normal variational equations along a non-equilibrium particular solution.
      We explicitly show that the connected component $G^0$ of the unit element of the differential
     Galois group of the normal variational equations along a simple rational solution is of the form $SL_2(\CC) \times SL_2(\CC)$ which is not Abelian.
       The obtained precise particular non-integrable result is extended to the entire orbit of the parameters by B\"{a}cklund transformations.

         It is tempting to utilize the Galois approach of non-integrability to other Sasano systems which have rational or algebraic particular solutions.
              We believe that the study the integrability of the  Sasano systems
       of dimension four and higher  will have influence on the development of the algebraic groups and Lie groups theory (classification of the algebraic subgroups of the group $SL_4(\CC)$),
            of the asymptotic and summability theory through the Stokes phenomenon and their applications to the mathematical physics.


   \vspace{1cm}

   {\bf Acknowledgments.}\,
   The author was partially supported by Grant KP-06-N 62/5  of the Bulgarian Fund
   "Scientific research".


 \vspace{1cm}
 {\bf Data availability statement.}\
   No new data were created or analysed during the current study.


\begin{small}
    
\end{small}

\end{document}